\documentclass[a4paper,11pt]{article}

\usepackage{jheppub} 

\usepackage{graphicx}
\usepackage{amssymb,amsmath}
\usepackage{bm}

\title{Complexity Functionals and Complexity Growth Limits in Continuous MERA Circuits}

\author[a]{J. Molina-Vilaplana,\note{Corresponding author.}}
\author[b,c]{A. del Campo}

\affiliation[a]{Universidad Polit\'ecnica de Cartagena,  C/Dr Fleming S/N. 30202 Cartagena, Spain}
\affiliation[b]{ University of Massachusetts,  Boston, MA 02125, USA}
\affiliation[c]{Theory Division, Los Alamos National Laboratory, MS-B213,  Los Alamos, NM 87545, USA}

\emailAdd{javi.molina@upct.es}
\emailAdd{adolfo.delcampo@umb.edu}

\abstract{Using the path integral associated to a cMERA tensor network, we provide an operational definition for the complexity of a cMERA circuit/state which is relevant to investigate the complexity of states in quantum field theory. In this framework, it is possible to explicitly establish the correspondence (Minimal) Complexity $=$ (Least) Action. Remarkably, it is also shown how the cMERA complexity action functional can be seen as the action of a Liouville field theory, thus establishing a connection with two dimensional quantum gravity. Concretely, the Liouville mode is identified with the variational parameter defining the cMERA circuit. The rate of complexity growth along the cMERA renormalization group flow  is obtained and shown to saturate limits which are in close resemblance to  the fundamental bounds to the speed of evolution in unitary quantum dynamics, known as quantum speed limits.  We also show that the complexity of a cMERA circuit measured through these complexity functionals, can be cast in terms of the variationally-optimized amount of left-right entanglement created along the cMERA renormalization flow. Our results suggest that the patterns of entanglement in states of a QFT could determine their dual gravitational descriptions through a principle of least complexity.}

\preprint{LA-UR-18-21985}

\begin{document} 
\maketitle
\flushbottom

\section{Introduction}
\label{Introduction}
In recent years, remarkable connections between gravitational physics and the patterns of entanglement in dual quantum states have
emerged \cite{ryu, VRK, epr_er, lashkari14, faulkner14}. These connections have been mainly addressed in the context of the AdS/CFT duality
\cite{malda98, gubser98, witten98}, a prominent example being the holographic formula for the entanglement entropy \cite{ryu}. In this context, the quantum complexity of a given state has been incorporated into the discussion very recently with two holographic proposals aiming to measure the quantum complexity of states in the boundary theory, namely, the Complexity=Volume (CV) conjecture \cite{susskind16, standford_susskind} and the Complexity=Action (CA) conjecture \cite{brown161, brown162}.

The complexity of a quantum state $|T\rangle$ is usually understood as the minimum number of gates needed to prepare the unitary transformation $U$ that obeys $|T\rangle=U|R\rangle$, where $|R\rangle$ is a simple reference state ~\cite{Aaronson16}. However, it is difficult to find a suitable definition of circuit complexity for states in a continuous theory such as a quantum field theory (QFT). The question is how to provide an operational definition of complexity based on smooth properties or observables. 
As argued in \cite{Chapman17}, the main challenges in providing this operational definition of complexity in the continuum include finding: i) suitable reference states, ii) a set allowed gates and its corresponding generators  that are simple (in the sense of \cite{susskind16, brown161, brown162, myers17, Chapman17}), iii) a regularization procedure to deal with ultraviolet divergences and iv) a measure of complexity. 
In \cite{Chapman17}, the choices for i)-iii) were inspired  by the continuous version of the Entanglement Renormalization tensor networks, cMERA \cite{Haegeman13, Takayanagi12}. For iv), a measure of the complexity of a quantum state  was introduced by integrating the Fisher information line element along a path from the reference state $|R\rangle$ to the target $|T\rangle$ associated to a realization of $U$. The unitary operators $U$ under consideration arise from iterating generators $K$ taken from some elementary set of Hermitian operators $\mathcal{K}$. Thus, unitaries $U$ can be represented as path ordered exponentials
\begin{align}
	\label{eq.defU}
	U(u) = \mathcal{P} \exp \left( - i \int_{u_i}^{u}  K(u') \, du'\right)  \, ,
\end{align}

where $u \in [u_i,u_f]$ parameterizes a path starting at $u_i$ and ending at $u_f$. The path-ordering $\mathcal{P}$ is required for non-commuting generators $K(u)$ and the interest lies on paths achieving 
\begin{equation}
|T\rangle \approx U(u_{f}) |R\rangle\, ,
\end{equation}

where $(\approx)$ indicates that states coincide for momenta below a cutoff $\Lambda$.  If the path is unrestricted, the unique unitarily invariant distance measure $d(|R\rangle, |T \rangle)=\arccos\left| \langle R| T \rangle \right| \leq \pi/2$ is obtained \cite{Chapman17}.  This is the familiar Bures length between  states $|R\rangle$ and $|T\rangle$ given by the (Fubini-Study) metric in  projective Hilbert space, see e.g. \cite{UC09}. By contrast, upon restricting the  generators $K(u)$,   a non-trivial notion of distance related to complexity may be obtained. In the proposal \cite{Chapman17}, the complexity ${\cal C}$ of a quantum state is defined as the minimal length (in terms of the Fisher metric) $\ell(|\Psi(u_f)\rangle) $ of a path running from $|\Psi(u_i)\rangle\approx |R\rangle$ to $\Psi(u_f)\approx |T\rangle$ driven by generators $K(u)$ in $\cal K$
\begin{equation}
	{\cal C} (|R\rangle,|T\rangle,{\cal K}, \Lambda) = \min_{K(u)} \ell(|\Psi(u)\rangle)\,.
\end{equation}

By definition, this measure of complexity $\cal C$ has properties of a distance function, inherited from the Fisher information metric. In addition, $\cal C$ remains well behaved even when $\cal K$ includes generators with unbounded norm (such as field operators). An interesting discussion of these ideas in the case of thermofield double states can be found in \cite{runqiu1}.

It is noteworthy that cMERA tensor networks appear in an alternative approach to the computation of the quantum complexity of a QFT state. In \cite{Takayanagi17a, Takayanagi17b, Czech18} an optimization procedure was introduced for Euclidean path-integrals that evaluate QFT wave functionals. This optimization is carried out by minimizing certain functional, which can be interpreted as a measure of the computational complexity, with respect to background metrics for the path-integrals. In two dimensional CFTs, the complexity functional is given by the Liouville action \cite{polyakov81, seiberg90}. The approach resembles a cMERA tensor network and the results were interpreted in terms of the AdS/CFT correspondence. Within this duality, and concretely under the CA conjecture, it has been recently recognized the significance of bounds on the complexification rates during a dynamical process. In \cite{brown161, brown162, runqiu2} authors found that in the context of black holes in AdS and thermofield double states, the speed of complexification of the boundary state saturates a bound analogous to that derived by Margolus and Levitin for unitary dynamics \cite{ML98}. The latter constitutes a fundamental  limit on the speed of evolution of any quantum system. More generally, the dynamics of physical processes is fundamentally constrained by so-called quantum speed limits, that set a bound to the rate of change of a measure of distance between states related by translation in time under the equations of motion of the system. Such bounds appear to be universal, and can be derived for both isolated \cite{MT45,Uhlmann92,ML98,LT09} and open \cite{QSLopen1,QSLopen2,QSLopen3} quantum systems as well as classical processes \cite{Shanahan18}. In particular, under unitary quantum dynamics, two-seminal results are known that bound the rate of change of the fidelity as a function of time in terms of the mean energy  (the aforementioned Margolus-Levitin speed limit \cite{ML98,LT09}) and via the energy fluctuations (Mandelstam-Tamm speed limit\cite{MT45,Uhlmann92}) of the system.  

In this work, we focus on the cMERA renormalization group flow for the ground state of free scalar theories for which some proposals for measuring complexity have shown to be explicitly computable \cite{Chapman17, myers17}. The cMERA tensor networks continuously project a reference state $|R\rangle$ into a final target $|T\rangle$ state, via successive application of the continuous equivalent of a single,  simple and fundamental, logic gate. This pictures a cMERA tensor network  for a free theory as a series circuit. It has been argued that the serial description is the one that applies naturally to black holes \cite{lloyd00}. We  show how the variationally optimized cMERA flows extremize the action functional appearing in the coherent state path integral representation for the circuit associated with the cMERA renormalization group (RG) flow. This path integral has been previously interpreted in terms of non-critical string in two dimensions \cite{molina16}. The extremal values of the action functional  account for the complexity of the cMERA circuit as measured in other proposals, see \cite{Chapman17}, and it is suggested to interpret the path integral as a complexity functional \cite{Takayanagi17a, Takayanagi17b, runqiu3}. In one dimensional theories, the complexity action functional for cMERA yields analogous results to the Liouville action of 2D Einstein gravity, which provides a connection between our proposal and those in \cite{Takayanagi17a, Takayanagi17b, Czech18}. To be concrete, the Liouville mode can be mapped to the variational parameter defining the cMERA circuit. In addition, understanding the cMERA RG flow as a dynamical process, we introduce Complexity Growth Limits (CGL) that strictly constrain the dynamics of the renormalization group flow implemented by the cMERA circuit.  In particular, we obtain CGL of Margolus-Levitin and Mandelstam-Tamm type that set upper bounds on the complexification rate of the circuit. We stress here that neither assumptions on the CA conjecture nor facts about the proposed holography/Tensor Network duality \cite{swingle12} have been considered in our derivations.

\section{Complexity in QFT and Entanglement Renormalization}
Tensor networks can be used as theoretical tools to characterize different features and properties of the wavefunction of a quantum many body system. The discrete version of an Entanglement Renormalization tensor network (MERA) \cite{vidal07} can be understood as a quantum circuit that builds a unitary operator that produces a target state from an initial set of decoupled qubits. Here we focus on the continuous version of MERA circuits (cMERA) \cite{Haegeman13} which allows to study the entanglement renormalization of states in continuous field theories.

\subsection{Continuous Entanglement Renormalization Circuits}
cMERA amounts to a real-space renormalization group procedure on the quantum state (instead of the Wilsonian RG scheme) that represents the wavefunction of the quantum system (usually in its ground state) at different length scales labeled by the parameter $u$. The renormalization scale parameter $u$ in cMERA  is usually taken to be in the interval $[u_{\scriptscriptstyle IR},u_{\scriptscriptstyle UV}] = (-\infty,0]$.  Here, $u_{\scriptscriptstyle UV} = u_{\epsilon}$ is the scale at the UV cutoff.  The real space UV cutoff is the short-wavelength cutoff $\epsilon$, and the corresponding momentum space UV cutoff is $\Lambda = 1/\epsilon$.  $u_{\scriptscriptstyle IR} = u_{\xi}$ is the scale in the IR limit, where $\xi$ is a long-wavelength correlation length.  In cMERA, the renormalized states $| \Psi(u) \rangle$  live all in the same Hilbert space $\mathcal{H}_{\Lambda}$, where $\Lambda$ is the UV-cutoff for the theory.  The state $|\Psi_{\scriptscriptstyle UV} \rangle = | \Psi(u_{\scriptscriptstyle UV}) \rangle$ is the state in the UV limit.  This may be the ground state of a many-particle system or the ground state of a quantum field theory.  The state
\begin{equation}
| \Omega \rangle = | \Psi_{\scriptscriptstyle IR} \rangle = | \Psi(u_{\scriptscriptstyle IR}) \rangle
\end{equation}	
is the unentangled reference state, with no entanglement between spatial regions, in the IR limit.  This state has zero entanglement entropy for any partition of space.  The cMERA-Hamiltonian
\begin{equation}
H_c(u) = K(u) + L
\end{equation}
generates translations along the cMERA parameter $u$.  $L$ is the (spatial) dilatation operator, governed by the scaling dimensions of the fields.  It can be understood as the ``free'' part of the cMERA-Hamiltonian, and does not depend on the cMERA parameter $u$.  The state $| \Omega \rangle$ is invariant under dilations, $L | \Omega \rangle = 0$. 

The term $K(u)$ in the cMERA-Hamiltonian is called the \emph{entangler} operator.  This is the ``interacting'' part of the cMERA-Hamiltonian.  The variational parameters of the cMERA formalism are incorporated into $K(u)$.  The cMERA unitary evolution operator $U(u_2,u_1)$ along the parameter $u$ is given by
\begin{equation}
U(u_2,u_1) = {\cal P} \exp \left[ -i \int_{u_1}^{u_2} du \ (K(u) + L ) \right]
\end{equation}
where ${\cal P} $ is the $u$-ordering operator.  States at different scales are thus related by $U(u_2,u_1)$.
In particular, we can represent the state $| \Psi(u) \rangle$ in terms of the IR state
\begin{equation}
| \Psi(u) \rangle = U(u,u_{\scriptscriptstyle IR}) | \Psi(u_{\scriptscriptstyle IR}) \rangle = U(u,u_{\scriptscriptstyle IR}) | \Omega \rangle\, .
\end{equation}

For most purposes it is convenient to define the cMERA ``interaction picture'' by the unitary transformation of states and operators
\begin{eqnarray}
| \Phi (u) \rangle &=&  \exp (i L u) | \Psi (u) \rangle\, ,\\ \nonumber
\tilde{O}(u) &=& \exp (i L u) \, O \exp (- i L u)\, .
\end{eqnarray}
In particular, the $u$-evolution in the interaction picture is determined by the unitary operator
\begin{equation}
\tilde{U}(u_2,u_1) = {\cal P} \exp \left[ -i \int_{u_1}^{u_2} du \, \tilde{K}(u) \right]\, ,
\end{equation}
with $\tilde{K}(u) = \exp (i L u) \, K(u) \exp (- i L u)$ the corresponding entangler in this picture. Thus, we may write
\begin{equation}
\label{cMERA_state}
| \Phi(u) \rangle = \tilde{U}(u,u_{\scriptscriptstyle IR}) | \Omega \rangle
= {\cal P} \exp \left[ -i \int_{u_{\scriptscriptstyle IR}}^{u} du' \, \tilde{K}(u') \right] | \Omega \rangle\, . 
\end{equation}
Starting in the IR state $| \Omega \rangle$, the interaction Hamiltonian $\tilde{K}(u)$ generates pairwise entanglement between modes of opposite momenta up to $| k| \le \Lambda e^u$, while for higher scales  only the dilatation operator $L$ acts. 
\\

In this work, we will consider the cMERA circuit associated to the ground state of a free scalar field theory in $(d+1)$-dimensions with a Hamiltonian in momentum space given by
\begin{align}
\label{hamilt_free}
H=\frac{1}{2}\int d^dk\,  [\pi(k)\pi(-k)+\omega_k^2\cdot \phi(k)\phi(-k)]\, ,
\end{align}

with $\omega_k = \sqrt{k^2 + m^2}$. It is convenient to express $\phi(k)$ and $\pi(k)$ in terms of creation and annihilation operators
\begin{align}
\phi(k)=\frac{a_k+a^{\dagger}_{-k}}{\sqrt{2\omega_k}}, \ \ \ \ \pi(k)=\sqrt{2\omega_k}\left(\frac{a_k-a^{\dagger}_{-k}}{2i}\right)\,,
\end{align}
with commutation relations  $[a_k,a^{\dagger}_p]=\delta^d(k-p)$. The $|\Omega\rangle$ unentangled (in real space) state is defined by


\begin{eqnarray}
\langle\Omega|\phi(k)\phi(k')|\Omega\rangle & =&\frac{1}{2M}\delta^{d}(k+k')\,, \\ \nonumber
\langle\Omega|\pi(k)\pi(k')|\Omega\rangle & = &\frac{M}{2}\delta^d(k+k')\,,
\end{eqnarray}

where $M=\sqrt{\Lambda^2 + m^2}$.  The cMERA variational optimization amounts to minimize the total energy $E = \langle \Phi(0) | H | \Phi(0) \rangle$, where $H$ is the Hamiltonian (\ref{hamilt_free}).

For this theory, the entangler in the interaction picture can be written in terms of creation and annihilation operators as the quadratic operator
\begin{equation}
\label{eq:disentangler}
\tilde{K}(u) = \frac{1}{2i}\int d^d k \, [\, g(k,u) \, a^{\dagger}_k a^{\dagger}_{-k} - g(k,u)^{*} \, a_k a_{-k} \,]\, ,
\end{equation}

where $ g(k,u) = g(u)\, \Gamma(|k| e^{-u}/\Lambda)$ with $\Gamma(x) =\theta(1-|x|)$ being a momentum cut -off function where $\theta$ is the step function. 
This is a Gaussian \textit{ansatz} which nomenclature is fully justified in \cite{Haegeman13, Takayanagi12, Cotler17} in which the state-dependent variational parameter to be determined is $g(k,u)$.  Working in the interaction picture, and noting that for the QFT vacuum $|0\rangle$,  $a_k|0\rangle = 0$, it is possible to write the cMERA state (\ref{cMERA_state}) as the squeezed state

	\begin{eqnarray}
	\label{cMERA_squeezed}
	|\Phi(u)\rangle = N\, \exp\Bigg[-\frac{1}{2} \int d^d k\, \left[\Phi(k,u)\,a^{\dagger}_k\, a^{\dagger}_{-k} - \Phi(k,u)^{*}\,a_k\, a_{-k}\right]\Bigg] |\Omega\rangle\, ,
	\end{eqnarray}

where 
\begin{equation}
\Phi(k,u) = \int_{-\infty}^{u} du' \, g(k,u')
\end{equation}

and the normalization constant reads
\begin{align}
N = \exp\left(-\frac{1}{2}\, \int_{|k|\leq \Lambda e^{u}} d^{d}k\, |\Phi(k,u)|^2\right)\, .
\end{align}

%
Thus, the cMERA RG-flow generates a sequence of scale-dependent Gaussian squeezed states $ |\Phi(u)\rangle$ that act as a set of scale-dependent Gaussian variational ansatz for the state $|\Phi_{\Lambda}\rangle\equiv|\Phi(u=0)\rangle$ by means of a variational parameter $\Phi(k,u)$ ($g(k,u)$).  In what follows, we will focus on the ground state of a $d=1$ scalar theory for which \cite{Haegeman13, Takayanagi12}
\begin{equation}
\Phi(k,u) = - \frac{1}{2} \log \frac{\omega_k}{M} \Big|_{k = \Lambda e^{u}}
= - \frac{1}{4} \log \frac{e^{2u} \Lambda^2 + m^2}{\Lambda^2 + m^2}\, ,
\end{equation}
that, differentiating with respect to $u$ gives 
\begin{equation}
g(u) = -\frac{1}{2}\frac{e^{2u} }{ (e^{2u}  + m^2/\Lambda^2)}\, .
\end{equation}

Finally, let us comment on the concept of \emph{simple} (non-orthogonalizing) quantum logic gates \cite{brown161, brown162, susskind16b, montero17}. It is said that a gate $\cal{G}$ operates simply on a state $|\Psi\rangle$ if $\langle \Psi|\mathcal{G}|\Psi\rangle\sim 1 -\mathcal{O}(\varepsilon)$ where $\varepsilon$ is the tolerance of the gate. This definition fits nicely with a cMERA tensor network pictured as a continuous quantum circuit.  Indeed, a cMERA tensor network implements a sequential quantum circuit on the reference state $|\Omega\rangle$ by sequential application of the \emph{simple} gate 
\begin{align}
\mathcal{G}_{\rm cMERA}(u)=e^{-i\, \delta\, \widetilde{K}(u)}\, ,
\end{align}

where $\delta$ amounts to an infinitesimal displacement along the renormalization direction $u$ that will be interpreted as the infinitesimal volume associated with the gate. This fulfills the criteria for it to be a \emph{simple} gate.  Interestingly,  it has been argued that any ``smooth" notion of complexity should be based upon simple gates which do not orthogonalize, i.e, gates that slightly change the wavefunction as the tolerance decreases \cite{montero17}.

\subsection{Complexity as cMERA Circuit Length}
In \cite{Chapman17}, authors have proposed to measure the complexity of a cMERA circuit as the the minimal length (in terms of the Fisher information metric) of a path running from the reference state $|\Omega\rangle$ to the target state $|\Phi_{\Lambda}\rangle$. This amounts to the real part of the quantum geometric tensor $g_{ij}$ introduced by Provost and Vallee \cite{PV80}. The later governs the (square of) Hilbert-Schmidt distance $d^2_{\rm HS}$ between quantum states $\Phi(\xi)$ and $\Phi(\xi+d\xi)$ that differ by an infinitesimal value of a set of parameters $\xi=(\xi_1,\xi_2,\dots,\xi_n)$, according to the expansion
\begin{eqnarray}
d^2_{\rm HS}\left[\Phi(\xi+d\xi),\, \Phi(\xi)\right]:=1-|\langle \Phi(\xi+d\xi)| \Phi(\xi)\rangle|^2 
=g_{ij}d\xi_id\xi_j\, ,
\end{eqnarray}
where the Fisher information metric reads
\begin{eqnarray}
g_{ij}={\rm Re}\langle \partial_i|\left(1-|\Phi \rangle\langle \Phi|\right)|\partial_j\Phi\rangle\, ,
\end{eqnarray}
with $\Phi=\Phi(u)$ for short. Thus, the proposal in \cite{Chapman17} can be cast as
\begin{align}
\mathcal{C}_{\rm cMERA}:=\int_{u_{\rm IR}}^{0}\, d_{\rm HS}\left[\Phi(u+du),\, \Phi(u)\right]\, .
\end{align}

Using the coherent state formulation of a cMERA circuit for a Gaussian theory, we first note that
\begin{eqnarray}
\langle \Phi(u+du)|\, \Phi(u)\rangle = e^{-\frac{\rm Vol}{2}\, \int d^d k\, |\Phi(k,u+du) - \Phi(k,u)|^2}  
=  e^{-\frac{\rm Vol}{2}\, \int d^d k\, |g(k,u)|^2\, du^2}\, ,
\end{eqnarray}
where  $\mathrm{Vol}\equiv \delta^d(0)$ is the (infinite) volume of the $d$-dimensional space $\mathbb{R}^{d}$. Thus, by definition we get
\begin{eqnarray}
d^2_{\rm HS}(u) = \mathcal{N}\, g_{uu}\, du^2 
=\mbox{Vol} \int_{|k|\leq \Lambda e^{u}}\, d^d k\, |g(u)|^2\, du^2\, ,
\end{eqnarray}

where $g_{uu}=|g(u)|^2$ and the normalization factor ${\cal N}$ is given by the volume of the effective phase space at length scale $u$:
\begin{align}
{\cal N}=\mbox{Vol}\cdot \int_{|k|\leq \Lambda e^{u}}\, d^d k\, .
\end{align}
Here we assume that $\Phi(u + du,k) \approx \Phi(u,k) + \partial_u\, \Phi(u,k)\, du$. In doing so, one may explicitly write
\begin{eqnarray}
\label{eq:circuit_lenght2}
\mathcal{C}_{\rm cMERA}^{(2)} = \int_{-\infty}^0\, du\, |g(u)|\, \sqrt{\mbox{Vol} \int_{|k|\leq \Lambda e^{u}}\, d^d k} 
=\sqrt{\mbox{Vol} \int_{|k|\leq \Lambda}\, d^d k\, \Phi_k^2} ,
\end{eqnarray}

where we have defined $\Phi_k \equiv \Phi(k,0)$ and the superscript $(2)$ relates to an interpretation of Eq.~(\ref{eq:circuit_lenght2}) as a $L^{2}$ norm. The result 
is independent of path reparameterizations. 

Alternatively, authors in \cite{Chapman17} also defined a $L^{1}$ norm version (Manhattan distance) of the circuit length given by
\begin{eqnarray}
\label{eq:circuit_lenght1}
\mathcal{C}_{\rm cMERA}^{(1)}= \mathrm{Vol} \int_{-\infty}^{0} du \, |g(u)|\, \int_{k < \Lambda e^u} d^d k\,  
= \,\mathrm{Vol} \, \int_{k \leq \Lambda} d^{d} k  \,  |\Phi_k| \, .
\end{eqnarray}

This definition amounts to disallowing different elementary gates in a circuit to act simultaneously and, according to our previous arguments, fits better as a measure for the sequential-single-simple gate structure of a cMERA circuit.  
\\

Explicit expressions for the complexity $\mathcal{C}_{\rm cMERA}^{(2)}$ of the ground state of the scalar theory\footnote{Results for $\mathcal{C}_{\rm cMERA}^{(2)}$ in free fermionic theories can be found in \cite{khan18}.} in arbitrary dimension $d$ can be found \cite{Chapman17} 
\begin{equation}
\frac{\Gamma \left(\frac{d}{2}+1\right)}{2 \pi ^{d/2} \text{Vol} \,\Lambda ^d} \, (\mathcal{C}_{\text{cMERA}}^{(2)}) ^2  = \frac{\Lambda^4 \, _2F_1\left(1,\frac{d+4}{4};\frac{d+8}{4};-\frac{\Lambda ^2}{m^2}\right){}^2}{4 (d+4)^2 m^4}\, .
\end{equation}
In the CFT case this simplifies to
\begin{equation}
\frac{\Gamma \left(\frac{d}{2}+1\right)}{2 \pi ^{d/2} \text{Vol} \, \Lambda ^d} \left( \mathcal{C}_{\rm cMERA}^{(2)} \right)^2 \biggr{|}_{m=0} = \frac{1}{4 d^2} \, 
\end{equation} 

which in the $d=1$ case yields
\begin{align}
\mathcal{C}_{\rm cMERA}^{(2)} = \sqrt{\mathrm{Vol}\cdot \Lambda}\, .
\end{align}

On the other hand, the complexity measure $\mathcal{C}_{\rm cMERA}^{(1)}$ for the ground state of the scalar theory in $d =1$ dimensions yields
\begin{align}
\label{circuit_length_1_CFT}
\mathcal{C}_{\rm cMERA}^{(1)}=\frac{\mathrm{Vol}\cdot \Lambda}{2}\, \Big[1- \frac{m}{2\Lambda}\left(\pi-2\arctan\left(\frac{m}{\Lambda}\right)\right)\Big]\, ,
\end{align}

which in the  the CFT limit ($m =0$) reduces to
\begin{align}
\label{eq:l1_massless}
\mathcal{C}_{\rm cMERA}^{(1)}=\frac{\mathrm{Vol}\cdot \Lambda}{2}\, .
\end{align}

\section{Complexity Action Functionals in cMERA}
Any complexity measure for a cMERA circuit must quantify the cost required to prepare the state $|\Phi_{\Lambda}\rangle$ from the specific reference state $|\Omega\rangle$ by applying the cMERA unitarity $U(0,u_{\rm IR})$ builded as the path ordered exponential of equation (\ref{cMERA_state}). In this Section we provide an alternative to the circuit length as a measure of complexity in cMERA circuits. The idea is to derive a new measure from a quantity able to encode the full RG-flow of a cMERA state and then evaluate it for concrete cMERA circuits.

 With this aim, we consider the coherent state path integral representation of the full cMERA RG flow associated to the ground state of a one dimensional free scalar theory\footnote{In \cite{molina16}, this path integral was also considered and the cMERA RG flow was formally interpreted in terms of the worldsheet action of a non-critical string in two dimensions. However, no physical content was assigned to this formal interpretation.}. This path integral is used as a recipe for mapping the Hamiltonian formulation of a cMERA circuit (\ref{cMERA_state}) into a Lagrangian one. We propose it as the quantity able to fully encode the RG-flow implemented by a cMERA circuit. To be explicit, we consider the amplitude
\begin{align}
\mathcal{Z}_{\rm cMERA}
=\langle \Phi_{\Lambda}\, |\mathcal{P}\, e^{-i\int_{u_{\rm IR}}^{0} du'\, \widetilde{K}(u')}    |\Omega\rangle\, .
\end{align}

Using standard techniques of coherent state path integrals which involve dividing the integration interval into $N$ infinitesimal subintervals and then inserting the resolution of identity 
\begin{align}
\int\, D\Phi\,  |\Phi(u)\rangle \langle \Phi(u)| = \mathbb{I}\, , 
\end{align}
at each subinterval limit, with $D\Phi$ being the gauge invariant Haar measure on SU(1,1)/ U(1), we obtain, after taking the limit $N\to \infty$

\begin{align}
\label{eq:cMERA_PI}
\mathcal{Z}_{\rm cMERA}=\int D\Phi\,D\Phi^{*}\, e^{i \mathcal{A}\left[ \Phi,\, \Phi^{*}\right] }\, .
\end{align}

Here, the quantum mechanical structure of the cMERA RG-flow reveals through the overlap of quantum states infinitesimally close in the $u$-direction.  The action functional in the path integral is given by 

\begin{align}
\label{eq:cMERA_action}
\mathcal{A}\left[ \Phi,\, \Phi^{*}\right]  =   -\int_{-\infty}^{0} du\, \Bigg(\mathcal{N}_{\Lambda}\, \Phi^{*}\partial\, \Phi + \widetilde{\mathcal{K}}\left[\Phi,\, \Phi^{*}\right] \Bigg) \, ,
\end{align}


where we used for short $\mathcal{N}_{\Lambda}\equiv {\rm Vol} \int_{ |k|\leq \Lambda} dk$, $\Phi \equiv \Phi(k,u)$, $\partial \equiv \partial_u$ and 
\begin{align}
\widetilde{\mathcal{K}}\left[\Phi,\, \Phi^{*} \right]  = \langle \Phi(u)|\widetilde{K}(u)|\Phi(u)\rangle\, .
\end{align}

Pursuing the analogy between the RG flow in a cMERA circuit and the time evolution in unitary quantum dynamics, 
the action functional is given by the sum of a geometric phase that arises from the Berry connection $\propto\Phi^{*}\partial\, \Phi $ and the analogue of the dynamical phase generated by $\widetilde{\mathcal{K}}\left[\Phi,\, \Phi^{*} \right]$, playing the role of the driving Hamiltonian.
\\

Noting that for the free scalar theory, $\Phi(k,u) \in \mathbb{R}$, the action $\mathcal{A}\left[ \Phi,\, \Phi^{*}\right]$ can be explicitly written as
\begin{align}
\label{eq:cMERA_action_explicit}
\mathcal{A}\left[ \Phi\right]  =  - 2\, {\rm Vol} \int_{|k|\leq \Lambda e^{u}}dk\, \int_{-\infty}^{0} du\, g(u)\, \Phi(k,u)  \, .
\end{align}

In this work, we propose that the complexity of a cMERA circuit can be measured in terms of $\mathcal{A}[\Phi]$. Concretely, we suggest that 
\begin{align}
\label{eq_CA_def}
\mathcal{C}_{\mathcal{A}}\left(|\Phi_{\Lambda}\rangle,\, |\Omega\rangle\right) := \mathcal{A}\left[ \Phi\right]_{\rm on-shell}\, ,
\end{align}

where \emph{on-shell} indicates that the action (\ref{eq:cMERA_action_explicit}) must be evaluated with the parameters $\Phi(k,u)$ obtained from the cMERA variational optimization. For simplicity, in the following we will use $\mathcal{C}_{\mathcal{A}}$ instead of $\mathcal{C}_{\mathcal{A}}\left(|\Phi_{\Lambda}\rangle,\, |\Omega\rangle\right)$, keeping in mind the explicit dependence on $|\Phi_{\Lambda}\rangle$ and the reference state $|\Omega\rangle$. 

Before moving forward, let us make few remarks on the explicit dependence of $\mathcal{C}_{\mathcal{A}}$ on the cMERA reference state. The IR reference state of a cMERA circuit is usually chosen as a topologically trivial state with no real-space entanglement. This choice has proven to yield correct results in the case of free scalar and free fermion theories \cite{Haegeman13, Takayanagi12}. In other words, with this choice, the variationally optimized cMERA parameters correctly reproduce correlators and entanglement entropies for the theories under consideration \cite{vidal17}. However, in \cite{ryu_top}, authors considered the cMERA RG flow in two dimensional free fermion theories with topologically non-trivial reference states. This was required in order to obtain suitable RG-flows (encoded by $g(k,u)$ and $\Phi(k,u)$) for describing two dimensional topological insulators. It was found that the non trivial topological entanglement of the initial reference state was reflected in the variational parameters of the cMERA circuit which showed non trivial modifications to those obtained with a trivial IR state. This result suggests that  $\mathcal{C}_{\mathcal{A}}$ might encode the choice of the initial reference state as the cMERA flow automatically captures its entanglement properties into the variational parameters that defines the full circuit.
\\

For the free scalar theory, in the gapless case with $m=0$ where $g(u)=-1/2$ and $\Phi(u)=-u/2$, the prescription in Eq. (\ref{eq_CA_def}) yields
\begin{align}
\label{eq:CA_massless}
\mathcal{C}_{\mathcal{A}} = -\frac{{\rm Vol}\cdot \Lambda}{2}\, \int_{-\infty}^{0}\, u\, e^{u}\, du = \frac{{\rm Vol}\cdot \Lambda}{2}\, .
\end{align}

that matches the result obtained by the circuit length $\mathcal{C}^{(1)}_{\rm cMERA}$ in (\ref{circuit_length_1_CFT}). 
One may also estimate the result for the massive scalar by assuming that $g(u)\approx-1/2$ and $\Phi(u)\approx-u/2$, for $u_{\rm IR}<u<0$ with $u_{\rm IR} \approx\log m/\Lambda$ to obtain
\begin{align}
\mathcal{C}_{\mathcal{A}}\approx \frac{{\rm Vol}\cdot \Lambda}{2}\Bigg[1 - \frac{m}{\Lambda}\left(1-\log \left(\frac{m}{\Lambda}\right)\right)\Bigg]\, .
\end{align}

\subsection{Complexity from Liouville Action}
The definitions and results  discussed above suggest that the cMERA action $\mathcal{A}[\Phi]$ is an efficient construction to encode the complexity of the RG flow. As an encoder of the target state complexity, it is worth pointing out that similar results can be obtained by considering the on-shell (in cMERA parlance) evaluation of the action
\begin{align}
\label{eq:Liouville_Phiz}
\mathcal{A}_L[\Phi]=\frac{1}{4}\int dx\, \int_{\epsilon}^{\infty}\, dz\, \Bigg[4 \left(\partial_{z}\Phi(z)\right)^2 + \Lambda^2\, e^{-4\Phi(z)} \Bigg]\, ,
\end{align}


with $z = \epsilon\, e^{-u}$ being a mere redefinition of the cMERA RG coordinate, $\epsilon = 1/\Lambda$ and ${\rm Vol} = \int dx$. 

%
For simplicity, we evaluate $\mathcal{A}_L[\Phi]$  in the massless case where 
\begin{eqnarray}
\label{eq:Phiz}
\Phi(z)= \frac{1}{2}\, \log\, \Lambda z\, ,
\end{eqnarray}

which being inserted into (\ref{eq:Liouville_Phiz}) yields,
\begin{align}
\mathcal{A}_L[\Phi(z)]_{\rm on-shell}=\frac{{\rm Vol}}{2\cdot \epsilon} = \frac{{\rm Vol}\cdot \Lambda}{2}\, .
\end{align}

This result suggests that  $\mathcal{A}_L[\Phi]$ can be regarded as an effective action for the cMERA path integral and our proposal for measuring the complexity of the target state $|\Phi_{\Lambda}\rangle$ can be expressed as 
\begin{align}
\label{eq:complexity_via_liouville}
\mathcal{C}_{\mathcal{A}}\equiv \mathcal{A}_L[\Phi]_{\rm on-shell}\, . 
\end{align}

What is most remarkable about $\mathcal{A}_L[\Phi(z)]$ in (\ref{eq:Liouville_Phiz}) is that, by defining $\Phi(z)$ in terms of a new field $\varphi_{L}(z)$ such that
\begin{align}
\label{eq:Phi_liouville}
\Phi(z)=\frac{1}{2}\, \log \Lambda - \frac{1}{2}\, \varphi_{L}(z)\, , 
\end{align}

one obtains
\begin{align}
\label{eq:Liouville_action}
\mathcal{A}_L[\varphi_{L}]=\frac{1}{4}\int dx\, \int_{\epsilon}^{\infty}\, dz\, \Bigg[\left(\partial_{z}\varphi_{ L}(z)\right)^2 +  e^{2\varphi_{ L}(z)} \Bigg]\, ,
\end{align}

that is the action of Liouville Field Theory (LFT) in its semi-classical limit \cite{polyakov81, seiberg90}. As a result, the Liouville field integral determines how the quantum correlations in a complex target state conspire to produce an emergent semi-classical description through the proposed connection between the cMERA variational parameter $\Phi(z)$ and the Liouville mode $\varphi_L$ in Eq. (\ref{eq:Phi_liouville}).


For a cMERA \textit{on-shell} evaluation of $\mathcal{A}_L[\varphi_{L}]$ we note that $\varphi_{L}(z)=-\log\,z$  in the massless case and thus
\begin{align}
\label{complexity_liouville}
\mathcal{A}_L[\varphi_{L}(z)]_{\rm on-shell}=\frac{{\rm Vol}}{2\cdot \epsilon} = \frac{{\rm Vol}\cdot \Lambda}{2}\, .
\end{align}

Interestingly, Einstein equations in two dimensions reduce to the Liouville's equation obtained from varying $\mathcal{A}_{L}[\varphi_L]$. In this sense, LFT provides a quantum theory of 2D-gravity that is called Liouville gravity and that is solved  by a metric given by
\begin{align}
ds^2 = e^{2 \varphi_{L}(z)}\, \left(dz^2 + dx^2\right)\, ,
\end{align}

with the boundary condition $ e^{2 \varphi_{L}(\epsilon)}=1/\epsilon^2$. The cMERA solution  $\varphi_{L}(z)=-\log\,z$ is therefore a solution of the  of Liouville's equation that imposes
\begin{align}
e^{2 \varphi_{L}(z)}=z^{-2}\, ,
\end{align}

leading  to the hyperbolic metric
\begin{align}
\label{hyperbolic_geom}
ds^2 =\frac{\left(dz^2 + dx^2\right)}{z^2} = du^2 +\Lambda^2\, e^{2u}\, dx^2\, .
\end{align}

Thus, one may say that the  geometry (\ref{hyperbolic_geom}) emerges from the tensor network circuit of least
complexity. Conversely,  solutions of the 2D gravity theory (\ref{eq:Liouville_action}) provide the correct  variational parameters that enable building suitable cMERA states/circuits with minimal complexity \cite{cheminassy16}. As a result, in this framework, it is possible to establish the correspondence {\it Least Action = Minimal Complexity}. These results can be consistently linked to those obtained in \cite{Takayanagi17a, Takayanagi17b, Czech18} and thus it would be sensible to consider (\ref{eq:cMERA_PI}) and its avatars as complexity path integrals or complexity functionals.

Let us illustrate this idea by considering the solution to the Liouville equation given by
\begin{eqnarray}
\label{tfd_geometry}
e^{2\, \varphi_L(z)} & =& \frac{\left(2 \pi /\beta\right)^2}{\sin^2\left(2 \pi\, z/\beta\right)}\, ,
\end{eqnarray}

with $0 < z <\beta/2$. This amounts to an asymptotically hyperbolic geometry (\ref{hyperbolic_geom}) close to the boundaries $z =\epsilon$ and  $z =\frac{\beta}{2}-\epsilon$ where $e^{2 \varphi_{L}(z)}=\epsilon^{-2}$. This geometry, after a suitable change of coordinates can be seen as the time slice of an eternal BTZ black hole \cite{maldacena01}, i.e an Einstein-Rosen bridge. By evaluating the complexity action functional (\ref{eq:Liouville_action}) one obtains
\begin{eqnarray}
\mathcal{A}_L[\varphi_L] &=& \frac{{\rm Vol}}{4}\, \int_{\epsilon}^{\frac{\beta}{2}-\epsilon}\, dz\, \left[(\partial \varphi_L)^2+ e^{2 \varphi_L}\right]= 2\, \Bigg[\frac{{\rm Vol}}{2\cdot \epsilon}\, \left(1- \frac{\pi^2\, \epsilon}{2 \beta}\right)\Bigg]\\ \nonumber
& = & 2\, \Bigg[\frac{{\rm Vol}\cdot \Lambda}{2}\, \left(1- \frac{\pi}{2}\, \frac{M}{\Lambda}\right)\Bigg]\, ,
\end{eqnarray}

with $M = \pi/\beta$. This is a remarkable result in view of the following problem: which is the cMERA circuit/state corresponding to this geometry? First, we note that the corresponding variational cMERA parameter $\Phi(z)$ should be given by
\begin{align}
\Phi_{\beta}(z)  =  \frac{1}{2}\log\, \Lambda\, \frac{\sin\left(2 \pi\, z/\beta\right)}{\left(2 \pi /\beta\right)}\, . 
\end{align}

Second, we note that at leading order
\begin{align}
\mathcal{C}_{\mathcal{A}}[\Phi_{\beta}] = 2\cdot \mathcal{C}_{\mathcal{A}}[\Phi_{M}]\, , 
\end{align}

where $\Phi_{M}$ amounts to the $\Phi(z)$ corresponding to a free scalar theory with mass $M=\pi/\beta$. According to this result, it is natural to think that the geometry (\ref{tfd_geometry}) encodes the complexity of a cMERA circuit consisting of two copies of the same circuit,
\begin{eqnarray}
|\Psi\rangle_{\beta} =  e^{-i \int_{u_{IR}}^0\, \widetilde{K}(u_1)\, du_1}\otimes  e^{-i \int_{u_{IR}}^0\, \widetilde{K}(u_2)\, du_2}\, |\Omega_\beta\rangle\, .
\end{eqnarray}

where $|\Omega_\beta\rangle$ is a pure state in the doubled Hilbert space. More details about this construction can be found in \cite{Takyanagi14}. 
\\

Let us briefly comment here on the possible higher dimensional generalization of $\mathcal{C}_{\mathcal{A}}$ which on general grounds simply reads as
\begin{align}
\mathcal{C}_{\mathcal{A}}[\Phi]:=-2\, {\rm Vol}\, \int_{|k| \leq \Lambda}d^{d}k\, \int_{-\infty}^{0}\, du\, \Phi(k,u)^{*}\, g(k,u)\, ,
\end{align}

with ${\rm Vol} = \delta^d(0)$ denoting the (infinite) volume of the $d$-dimensional space $\mathbb{R}^{d}$. In cMERA RG-flows in higher dimensions (see for instance \cite{Cotler17, ryu_top, khan18}) the variational parameters $\Phi(k,u)$ and $g(k,u)$ are such that this complexity functional naturally produces leading divergent terms 
\begin{align}
\mathcal{C}_A \sim {\rm Vol} \cdot \Lambda^d\, .
\end{align}

In \cite{Takayanagi17a, Takayanagi17b}, it has been argued that an effective Liouville-like complexity functional that could, upon minimization,  account for this kind of leading divergent behaviour,  is given by
\begin{align}
\mathcal{A}_{L}^{(d)}[\varphi_L]\propto\int dx^d\, dz\, e^{(d-1)\varphi_L}\left[(\partial_{z}\varphi_L)^2 + e^{2\varphi_L}\right]\, .
\end{align}
Therefore, the minimization of $\mathcal{A}_{L}^{(d)}[\varphi_L]$ yields the hyperbolic space $\mathbb{H}^{d+1}$ which amounts to the time slice of pure AdS$_{d+2}$.

\subsection{Perturbed cMERA circuits and Vertex Operators in LFT.}
From part of our previous analysis, the complexity of a cMERA circuit (e.g, the circuit for the ground state of a massless scalar theory), measured by a Liouville action functional such as (\ref{eq:Liouville_action}), is related  to the choice of the {\it simple} cMERA gates of the form
\begin{align}
\mathcal{G}_{\rm cMERA}\sim e^{-i\delta_{\rm Vol}\, \Phi(z)\, \mathcal{O}}\, ,
\end{align}

where $\mathcal{O}$ is an operator of the $\mathfrak{su}(1,1)$ algebra. One may wonder what would be the effect of perturbing or deforming these gates at some points of the renormalization scale direction $z$. Here, we provide an argument based on the properties of correlation functions in Liouville field theory. To this end,  we first assume  the perturbed gates to be located at a finite number of points $\left\lbrace z_i \right\rbrace_{i=1\cdots m}$. Second, we impose that the perturbation  affects the shape of $\Phi(z)$ only in a small neighborhood $\varepsilon_i$ around each of the points $z_i$. From the connection between the  $\Phi(z)$ in the cMERA circuit and the Liouville mode $\varphi_L(z)$ appearing in $\mathcal{A}_L$, one can infer that a suitable ansatz for the deformed $\Phi(z)$ would come from a sensible choice for a deformed $\varphi_L(z)$ in small neighborhood of $z_i$. 

Thus, we consider the LFT correlation functions 
\begin{align}
\label{Liouville_vertex}
\left\langle \prod_{i}\, e^{\alpha_i\, \varphi_L(z_i)}\right\rangle = \int\, \mathcal{D}\varphi_{ L}\, e^{\mathcal{A}_L[\varphi_L]}\,  \prod_{i}\, e^{\alpha_i\, \varphi_L(z_i)}\, , 
\end{align}

where operators $V(\alpha_i,z_i)=e^{\alpha_i\, \varphi_L(z_i)}$ are known as vertex operators. For values of $\alpha_i$ which do not exceed some bounds (see \cite{seiberg90})  the functional integral (\ref{Liouville_vertex}) is dominated by configurations of $\varphi_L$ satisfying
\begin{eqnarray}
\label{eq:vertex_Liouville}
\varphi_L(z)&=&-\log\, z\, \quad {\rm when}\, \quad  |z-z_i| \gg \varepsilon_i\, \\ \nonumber
\varphi_L(z)&=&-2\alpha_i\log\, |z - z_i| + \varphi_i + \cdots\, \quad {\rm at}\, \quad  |z-z_i| \leq \varepsilon_i \, .
\end{eqnarray}

We notice that this is precisely the kind of deformation on $\varphi_{ L}$ and hence on a $\Phi(z)$-cMERA gate that we were looking for. Thus, we will interpret the action functional in (\ref{Liouville_vertex})
\begin{align}
\mathcal{A}_L[\varphi_L,\alpha_i]=\mathcal{A}_L[\varphi_L] + \sum_i\, \alpha_i\, \varphi_L(z_i)\, ,
\end{align} 

evaluated at (\ref{eq:vertex_Liouville}), as the action functional which measures the complexity of the corresponding deformed cMERA circuit. A direct evaluation leads to $\log|z_i - z_i|$ divergences arising in the vicinity of the insertion points $z_i$. A proper regularization is given (see \cite{seiberg90} for details) by,
\begin{align}
\label{eq:reg_Liouville_Vertex}
\mathcal{A}_L[\varphi_L,\alpha_i]=\mathcal{A}_L[\varphi_L]_{/ \cup \varepsilon_i} + \sum_i\Big(\alpha_i\, \varphi_i - 2\alpha_i^2\log\, \varepsilon_i^2\Big)\, , 
\end{align}

where $/ \cup\, \varepsilon_i$ means that the evaluation excludes the vicinity of the insertion points $z_i$. Here, we are not interested in explicit results on the evaluation of (\ref{eq:reg_Liouville_Vertex}) but only in noticing that the vertex operator insertion model for deforming the cMERA circuit leads to a complexity increment given by
\begin{align}
\Delta\, \mathcal{C}_{\mathcal{A}}(\alpha_i) =  \sum_i\Big(\alpha_i\, \varphi_i - 2\alpha_i^2\log\, \varepsilon_i^2\Big)\, .
\end{align}

While the exact value for $\Delta\, \mathcal{C}_{\mathcal{A}}(\alpha_i)$ in the case of an arbitrary number of insertions  is not known explicitly, results for some number of concrete settings can be found in \cite{harlow11}. 

\paragraph{Complexity Functionals in Interacting Field Theories.}
Let us finish this Section by commenting on the possible validity of our proposal in the case of interacting field theories. Despite the cMERA circuit lacks the generality of its discrete version, first steps to systematically build a variational approximation to deal with interacting theories has recently been proposed \cite{Cotler17, Cotler18a, Cotler18b}. Here we briefly summarize this approach and comment on its implications to evaluate the complexity of target states in interacting theories.

Due to the multipartite structure of quantum correlations expected for the ground states of interacting QFT's, a proposal to adapt cMERA  to this situation amounts to expand the entangler $\widetilde{K}(u)$ as \cite{Cotler17}
\begin{align}
\label{entangler1}
\widetilde{K}(u) = &\frac{1}{2 i} \int d^d k_1 \,\left(g_{1,0}(k_1; u) \, a_k^\dagger - \text{h.c.}\right) \nonumber \\
&+ \frac{1}{2 i} \int d^d k_1 \int d^d k_2 \, \left(g_{2,0}(k_1, k_2; u) \, a_{k_1}^\dagger a_{k_2}^\dagger + g_{1,1}(k_1, k_2; u) \, a_{k_1}^\dagger a_{k_2} - \text{h.c.} \right) \nonumber \\
& + \frac{1}{2 i} \int d^d k_1 \int d^d k_2 \int d^d k_3 \, \left(g_{3,0}(k_1,k_2,k_3; u) \, a_{k_1}^\dagger a_{k_2}^\dagger a_{k_3}^\dagger + \cdots - \text{h.c.} \right) \nonumber \\
& + \,\, \cdots
\end{align}

where now the optimization runs over the variational functions $g_{i,j}$. It is clear that for free theories, $\widetilde{K}(u)$ in Eq. \!(\ref{entangler1}) is truncated after the first two lines to yield the cMERA circuit defined by Eq. \!(\ref{eq:disentangler})\footnote{Namely, in Eq. \!(\ref{eq:disentangler}) we set all $g_{i,j}=0$ except $g_{2,0}(k,-k;u)\equiv g(k,u)$.}. The physical interpretation of this expansion is the following:  with the quadratic terms we capture the entanglement structure and thus the complexity of our target state due to pairwise correlations between modes, the cubic terms account for tripartite correlations between modes and so on.  Thus, truncating Eq. \!(\ref{entangler1}) at order $n$ means the truncated ansatz account at most for $n$-partite correlations at each scale.

Regarding our proposal to measure complexity, going beyond quadratic order in the entangler and thus building a path integral representation for this generalized cMERA circuit,  one faces  a major difficulty: to efficiently compute expectation values with a non-quadratic ansatz $|\Phi(u)\rangle$ built as a path ordered exponential (see (\ref{cMERA_state})). However, as the entangler $\widetilde{K}(u)$ lies inside a path-ordered exponential, a possible approach to address this problem is to perform perturbation theory on the non-quadratic terms.  Despite this approximation being perturbative from the point of view of the original entangler, the resulting variationally optimized ansatz resums  infinite classes of Feynman diagrams, allowing to access in principle, non-perturbative physics \cite{PostGaussian1, PostGaussian2, PostGaussian3, PostGaussian5, PostGaussian6}. Very recent developments on cMERA circuits for weakly interacting theories have been reported in \cite{Cotler18a, Cotler18b}.  There, authors use a perturbation theory version of a cMERA quantum circuit to construct a \emph{local} entangler  $\widetilde{K}(u)$ such that the corresponding cMERA state agrees with the 1-loop UV ground state of the $\phi^4$ interacting scalar theory. Due to the locality of the entangler, in principle it would be possible to retain the notion of simple cMERA gates and, in a future work, study perturbative corrections to the complexity of a target state by means of cMERA circuits.

\section{Complexity and Entanglement in cMERA}
\label{section:entanglement}
In this Section we establish explicit connections between the complexity of a target cMERA state and its entanglement structure, conveniently encoded in the variational parameters defining its associated cMERA circuit. First, we would like to {\bf recover the}  connection between the Liouville mode $\varphi_L$ and the entanglement entropy of the ground state of the scalar theory noticed by \cite{Takayanagi17b, Czech18}. Here, the cMERA circuit construction is used to derive it. To this end, we note that a cMERA circuit is builded by taking $|\Omega\rangle$,  the totally factorized reference state, and then adding a variationally prescribed amount of entanglement between neighboring regions at scale $u$. This is precisely the pairwise entanglement created between modes with opposite momenta $|k|=\Lambda\, e^{u}$. To be precise, the entanglement between neighboring ``sites" at a scale $u = -\log\, \Lambda\, L$ represents, in the UV, the entanglement between regions of size $L/\epsilon$ and the rest of the system. Remarkably, it is the variational parameter $\Phi(z)$ 
which controls the precise amount of this scale-dependent pairwise entanglement \cite{molina15, Cotler17}. In the massless scalar theory the ground state entanglement for an interval $A$ of length $L$ is given by
\begin{align}
S_A[L]=\frac{c}{6}\, \log \frac{L}{\epsilon} = \frac{c}{6}\, \log \Lambda\, L \, ,
\end{align}

where $c=1$ is the central charge of the CFT. Recalling (\ref{eq:Phiz}), it follows  that 
\begin{align}
\frac{6}{c}\, S[z] = 2\, \Phi(z)=\log\Lambda - \varphi_L[z]\, ,
\end{align}


from which it is possible to see how the Liouville-like equation of motion
\begin{align}
\partial\Bigg(\frac{6}{c}\, S[z]\Bigg) = \Lambda\, e^{-\frac{6}{c}\, S[z]}\, 
\end{align} 
dictates the entanglement structure of the quantum state under consideration. In other words, the Liouville mode profile that measures the complexity of the target state through $\mathcal{A}_L$ is shown to be directly related with the entanglement structure of the target state.
\\

In addition, we also provide an explicit connection between the complexity of a cMERA circuit, measured through the circuit length, for the ground state of a free scalar theory, and the quantum fluctuations created along the cMERA renormalization group flow of the wavefunction. As commented above,  the  amount of scale-dependent entanglement that must be added to the reference state $|\Omega\rangle$ to create $|\Phi_{\Lambda}\rangle$ is given in terms of the left-right entanglement (LREE) between modes with opposite momenta appearing in (\ref{cMERA_squeezed}). This can be measured by the von Neumann entropy of the reduced density matrix obtained by tracing out the (for instance) left moving modes \cite{molina15}. In a free theory where all modes are decoupled one may focus on
\begin{equation}
\rho_{k}^{R}(u) = (1- \gamma_{k}(u)) \, {\rm diag}\left(1, \gamma_k(u), \gamma_k(u)^2,\gamma_k(u)^3\, \cdots \right) \, ,
\end{equation}
where
\begin{equation}
\gamma_{k}(u) = \tanh^2 \Phi(k,u)\, .
\end{equation}

In \cite{molina15} the rate at wich LREE ($S^{\pm}$) is created along the cMERA RG-flow was obtained, yielding
\begin{equation}
\partial\, S^{\pm}_k(u)\approx 2\, g(k,u)\, .
\end{equation}

This explicitly relates the rate of LREE generation with the variational strength of the \textit{entangling} operation $g(k,u)$. As a result, the total amount of LREE created along a cMERA RG-flow is given by

\begin{eqnarray}
S^{\pm} \approx 2 \int_{k\leq \Lambda} d k \, \int_{-\infty}^{0} du\,  g(k,u) 
= 2 \int_{k\leq \Lambda} d k \, \Phi_k\, .
\end{eqnarray}

According to the expressions given above and recalling Eqs.(\ref{eq:circuit_lenght2}) and (\ref{eq:circuit_lenght1}), it is possible to establish 
\begin{align}
\mathcal{C}^{(2)}_{\rm cMERA}=\frac{1}{2}\, \int_{-\infty}^{0} du\, |\partial S^{\pm}_k(u)|\, \sqrt{\mbox{Vol} \int_{|k|\leq \Lambda e^{u}}\, d k}\, ,
\end{align}

and
\begin{align}
\label{eq:complexity1_LREE}
\mathcal{C}^{(1)}_{\rm cMERA}=\frac{\mathrm{Vol}}{2} \, S^{\pm}\, .
\end{align}

These results show that the complexity of the cMERA state measured through  its circuit length, can be cast in terms of the total amount of the scale-dependent entanglement created along the  renormalization of the quantum state. Thus, the complexity of the state is directly related to the depth of the renormalization group transformation that is needed to project the generally highly entangled UV state to a direct product state in the IR. For gapless states, this entanglement persistently spreads out to larger length scales. It is noteworthy  that the total amount of LREE is given in terms of the Liouville-like scalar $\Phi(z)$ that, as shown above, it is directly related with the central charge of the  theory under consideration. Indeed, regarding Eq.(\ref{eq:complexity_via_liouville}), this may suggest that the patterns of entanglement in states of a QFT could determine their dual gravitational descriptions through a principle of least complexity.

\section{Complexity Growth Rates}
Given the structure of the cMERA RG-flow encoded in the action $\mathcal{A}[\Phi]$ of Eq.(\ref{eq:cMERA_action}) (i.e, a Lagrangian representation of the RG - dynamics along the renormalization direction under the action of the entanglement Hamiltonian $\widetilde{K}(u)$), one might wonder if the aforementioned complexity measure $\mathcal{C}_{\mathcal{A}}$ is fundamentally constrained in some sense. We next show that this is indeed the case, by introducing a bound to the complexification dynamics inherent to any cMERA circuit that we shall refer to as Complexity Growth Limit (CGL). To this end, let us define the complexity of the intermediate state $|\Phi(u)\rangle$, $\mathcal{C}_{\mathcal{A}}(u)\equiv\mathcal{C}_{\mathcal{A}}\left(|\Phi(u), |\Omega\rangle\right)$  given by

\begin{align}
\mathcal{C}_{\mathcal{A}}(u)=-2\, {\rm Vol}\, \int_{k \leq \Lambda}dk\, \int_{-\infty}^{u}\, du'\, \Phi(k,u')^{*}\, g(k,u')\, ,
\end{align}

from which, the rate of growth reads 
\begin{align}
\frac{d\, \mathcal{C}_{\mathcal{A}}(u)}{du} =-2\, {\rm Vol}\, \int_{k \leq \Lambda}dk\,  \Phi(k,u)^{*}\, g(k,u)\, . 
\end{align}

On the other hand, regarding Eqs.(\ref{eq:disentangler}) and (\ref{cMERA_squeezed}), a straightforward calculation gives
\begin{eqnarray}
\langle \widetilde{K}(u) \rangle  = \langle \Phi(u)| \widetilde{K}(u)|\Phi(u) \rangle 
=  {\mathrm{Vol}}\, \int_{k \leq \Lambda} d k \,  \Phi(k,u)^{*}\, g(k,u)\, ,
\end{eqnarray} 

thus allowing to write,
\begin{align}
\frac{d\, \mathcal{C}_{\mathcal{A}}(u)}{du} = -2\, \langle \widetilde{K}(u) \rangle\, .
\end{align}
Given  that $u\in(-\infty,0]$, we find upon integration,
\begin{align}
\label{QCGR_ML}
u_f-u_i= \frac{\mathcal{C}_{\mathcal{A}}(u_f)-\mathcal{C}_{\mathcal{A}}(u_i)}{2\,  \widetilde{\mathbf{K}} }\, ,
\end{align}
which provides a CGL of the Margolus-Levitin type, by analogy with the quantum speed limit known for unitary quantum dynamics  \cite{ML98,LT09}. In particular, (twice) the time-averaged mean value of the entanglement Hamiltonian 
\begin{align}
\label{}
\widetilde{\mathbf{K}}=\frac{1}{u_f-u_i}\int_{u_i}^{u_f} \langle \widetilde{K}(u) \rangle du\, ,
\end{align}
provides the  CGL, in close correspondence with the time-averaged mean energy in the case of unitary dynamics of driven systems \cite{DL013}. 
In our analogy, $u$ plays the role of time and the entanglement Hamiltonian that of the driving Hamiltonian generating the time evolution.

However, Equation(\ref{QCGR_ML}) is an equality as opposed to a lower bound on the required $u$-shift. This reflects the optimality of the  entanglement Hamiltonian for the generation of complexity along the cMERA circuit. Physically this means that for a target state with a fixed complexity measured by $\mathcal{C}_A$, from the reference state $|\Omega\rangle$, a cMERA circuit creates complexity (measured by $\mathcal{C}_A$) at a maximum rate, which amounts to an optimal circuit with minimum depth $u$. We would like to emphasize this last point. One might wonder if a true minimal complexity can be achieved using the tensor network path integral. As proving optimality in quantum field theory is a difficult task at this moment, the saturation of fundamental limits in the form of our proposed CGL, seems to be a useful tool for 
assessing circuit optimality.

\subparagraph*{A Complexity Growth Rate from Circuit Length.}

It is also  possible to derive a CGL for the rate of complexity when the circuit lengths $\mathcal{C}_{\rm cMERA}^{(1)}$  and $\mathcal{C}_{\rm cMERA}^{(2)} $ are  considered as the measure for the complexity of the target state.  We wonder how these complexity measures, based on the distance between states that are related by the evolution under the action of $\widetilde{K}(u)$, are constrained.  Proceeding in a similar way as before, we first consider the complexity of the cMERA state $|\Phi(u)\rangle$ in terms of the circuit length $\mathcal{C}_{\rm cMERA}^{(2)}$
\begin{align}
\mathcal{C}_{\rm cMERA}^{(2)}(u)= \int_{-\infty}^{u}\, |g(u')|\, du'\, \sqrt{\mbox{Vol} \int_{|k|\leq \Lambda e^{u}}\, d^d k}\, .
\end{align}
Taking the derivative with respect to $u$ yields
\begin{eqnarray}
\frac{d\, \mathcal{C}_{\rm cMERA}^{(2)}(u)  }{du} =  |g(u)|\,  \sqrt{\mbox{Vol} \int_{|k|\leq \Lambda e^{u}}\, d^d k}
=\sqrt{\mathcal{N}}\, |g(u)|=\Delta \tilde{K}(u)\, ,
\end{eqnarray}

where  we used that $\mathcal{N}\, |g(u)| \equiv \Delta \tilde{K}(u)$ with \cite{Takayanagi12}
\begin{align}
\Delta \tilde{K}(u)^2=\langle \Phi(u)| \tilde{K}^2(u)|\Phi(u)\rangle - \langle \Phi(u)| \tilde{K}(u)| \Phi(u)\rangle^2\, .
\end{align}
We can readily integrate this differential equation to rewrite it in a form resembling a QSL,
\begin{eqnarray}
\label{QCGR}
u_f-u_i=\frac{\mathcal{C}_{\rm cMERA}^{(2)}(u_f)-\mathcal{C}_{\rm cMERA}^{(2)}(u_i)}{\Delta\widetilde{\mathbf{K}}}\, , 
\end{eqnarray}
where we have introduced the averaged dispersion of the entanglement Hamiltonian
\begin{eqnarray}
\Delta\widetilde{\mathbf{K}}=\frac{1}{u_f-u_i}\int_{u_i}^{u_f}\Delta \tilde{K}(u)du\, .
\end{eqnarray}
Therefore, the rate of change of the complexity is exactly set by the averaged fluctuations of the entanglement Hamiltonian along the cMERA RG flow. This result is analogous to the Mandelstam-Tamm bound for unitary quantum dynamics  \cite{MT45}  generated by a driving Hamiltonian. In particular,  it  provides a CGL equivalent to that for the rate of change of the fidelity between a quantum state and its time-evolved state via the time-average of the energy fluctuations \cite{AA90,Uhlmann92}. As before, in this analogy, $u$ plays the role of time and the entanglement Hamiltonian that of the driving Hamiltonian generating the time evolution. Likewise, Equation (\ref{QCGR}) is an equality and not only a lower bound on the required $u$-shift. Again, this is a signature of the optimality of the entanglement Hamiltonian for the generation of complexity in analogy with the quantum brachistochrone problem that concerns the time-optimal evolution, saturating the Mandelstam-Tamm bound \cite{Carlini06}.

A similar Mandelstam-Tamm like CGL can be derived using the circuit length $\mathcal{C}_{\rm cMERA}^{(1)}$ by noticing that
\begin{eqnarray}
\frac{d\, \mathcal{C}_{\rm cMERA}^{(1)}(u)  }{du} = \mbox{Vol}\,  |g(u)|\,   \int_{|k|\leq \Lambda e^{u}}\, d^d k= \mathcal{N}\, |g(u)|= \sqrt{\mathcal{N}}\Delta \tilde{K}(u)\, .
\end{eqnarray}

We remark here that the Margolus-Levitin CGL can not be directly obtained from the complexity measure $\mathcal{C}_{\rm cMERA}^{(1)}$.

\subsection{Complexodynamics in Time Dependent Settings}
At this point, it is natural to ask whether our proposal to measure the complexity and complexity rates by means of action functionals of the type in Eq. (\ref{eq:Liouville_action}) can be applied to settings in which the states are subjected to real time evolutions 
\begin{align}
|\Phi_{\Lambda}(t)\rangle=e^{-iHt}\, |\Phi_{\Lambda}\rangle\, ,
\end{align}

 with generic Hamiltonians denoted by $H$. In general, a unique complexity measure for unitarities $U(t)= e^{-iHt}$ can not be defined a priori. As we have discussed, any operational definition of complexity relies on having a set allowed gates and its corresponding generators that must be simple, something that is not guaranteed in generic situations.  Namely, by measuring the complexity with a gate set that depends directly on $H$, it is expected that the complexity will be a function of both the Hamiltonian and the state. In other words, by changing  $H$, one must change the way to measure the complexity \cite{brown162}. 
 \\
 
 Here, we show that a formalism developed in \cite{Takayanagi12, Takyanagi14} to deal with time-dependent states in cMERA can overcome these difficulties. The formalism addresses time-dependent states generated by  a quantum quench. Quantum quenches induce a time evolution on an initial state $|\Phi_{\Lambda}\rangle$ due to a sudden change of the Hamiltonian. For instance,  the time evolved states can arise due to an instantaneous change of the mass parameter in the scalar theory. Authors in \cite{Takayanagi12, Takyanagi14} build up a cMERA circuit that maps the initial reference state $|\Omega\rangle$ onto the target state $|\Phi_{\Lambda}(t)\rangle$ by solving a time dependent variational problem such that
\begin{align}
\label{eq:cMERA_time}
|\Phi_{\Lambda}(t)\rangle=  {\cal P} \exp \left[ -i \int_{u_{\scriptscriptstyle IR}}^{0} du\, \widetilde{K}(u;t) \right] | \Omega \rangle\,  ,
\end{align}

where, remarkably, $\widetilde{K}(u;t)$ is a cMERA entangler Hamiltonian with the same structure as (\ref{eq:disentangler}) while  the variational parameter $g(k,u;t)$ is now time dependent. This is relevant  as it provides us with a set of \emph{simple} cMERA gates that can be used to evaluate the complexity of the time dependent state $|\Phi_{\Lambda}(t)\rangle$ by directly evaluating the complexity functionals presented above.
\\

Following \cite{Takyanagi14}, we focus on a special process where the mass parameter is suddenly shifted from a non-zero value $\Delta m$ to zero such that  $\Delta m \sim 1/\mathcal{J}$. We note that the quantum quench has no effect for $\Delta t / \mathcal{J} \geq 1$. As a result, quantities such as entanglement entropy, grow with a concrete time dependence for $\Delta t/\mathcal{J} \leq 1$ and then saturate at a final value for $\Delta t \geq \mathcal{J}$ \cite{calabrese05}. In order to present our result, let us introduce the following notation: for the case of the massless scalar we recall the definition of the cMERA-Liouville mode $\varphi_L(z)$ via
\begin{align}
e^{\varphi_L(z)}=\chi(z)= z^{-1}\,  ,
\end{align} 

with $\chi(z)= 2\,\partial_z\, \Phi(z)$. In \cite{Takyanagi14} it is shown that for the mass quench in the scalar theory
\begin{align}
\chi(z, t)\approx\frac{1}{z}\, \Bigg(1 + \frac{A\, k t+ B\, k\mathcal{J}}{\sinh(k \mathcal{J}/2)}\Bigg)\, ,
\end{align}

where $k \sim 1/z$ and $A$ and $B$ are order one positive constants. As the quench has effect only for modes with $k\,\mathcal{J} \ll 1$, we approximate 
\begin{align}
e^{\varphi_L(z,t)}=\chi(z,t)\approx\frac{1}{z}\, \Bigg(1 + \frac{t}{ \mathcal{J}}\Bigg)\, ,
\end{align}

or equivalently 
\begin{align}
\varphi_L(z,t)= -\log z + \log \Bigg(1 + \frac{t}{ \mathcal{J}}\Bigg) = \varphi_{ L}(z) + \log \Bigg(1 + \frac{t}{ \mathcal{J}}\Bigg)\, .
\end{align}

Finally, we proceed to evaluate the complexity $\mathcal{C}_{\mathcal{A}}(t)\equiv\mathcal{C}_{\mathcal{A}}(|\Phi_{\Lambda}(t)\rangle,|\Omega\rangle)$ by means of the Liouville complexity functional as
\begin{align}
\mathcal{C}_{\mathcal{A}}(t)=\mathcal{A}_L[\varphi_L(z,t)] = \frac{1}{4}\int dx\, \int_{\epsilon}^{\mathcal{J}}\, dz\, \Bigg[\left(\partial_{z}\varphi_{ L}(z,t)\right)^2 +  e^{2\varphi_{ L}(z,t)} \Bigg]\,  ,
\end{align}

which yields
\begin{align}
\mathcal{C}_{\mathcal{A}}(t)\approx\frac{{\rm Vol}}{4}\, \int_{\epsilon}^{\mathcal{J}}\, \frac{dz}{z^2}\Big(2 + \frac{2\, t}{\mathcal{J}}\Big) = \underbrace{\frac{{\rm Vol}\cdot \Lambda}{2}\, \Bigg(1-\frac{\Delta m}{\Lambda}\Bigg)}_{ \mathcal{C}_{\mathcal{A}}(0)}\, \Bigg(1 + \frac{ t}{\mathcal{J}}\Bigg) 
 =  \mathcal{C}_{\mathcal{A}}(0)\, \Bigg(1 + \frac{ t}{\mathcal{J}}\Bigg)\, .
\end{align}

As a result, we see that the complexity of the time evolved state $|\Phi_{\Lambda}(t)\rangle$ grows linearly in time until it saturates for times $t \geq \mathcal{J}$. During this period, the rate for the complexity growth amounts to a constant which is given by
\begin{align}
\frac{d\,  \mathcal{C}_{\mathcal{A}}(t)}{dt} = \mathcal{C}_{\mathcal{A}}(0)\, \Delta m\, \approx\frac{{\rm Vol}\cdot \Lambda}{2}\, \Delta m  + \mathcal{O}(\Delta m/\Lambda)\, .
\end{align}

It is worth to compare these results with estimations on the complexity growth measured in terms of the circuit length $\mathcal{C}_{\rm cMERA}^{(1)}$. As shown in Section \ref{section:entanglement}, $\mathcal{C}_{\rm cMERA}^{(1)}$ can be cast in terms of the total amount of LREE created along the cMERA RG-flow. In addition, it was shown in \cite{Takyanagi14} that the growth of the entanglement entropy over time $\Delta S_A(t)$ for a subsystem $A$ that amounts to the half space is 
\begin{align}
\Delta S_A(t)\equiv \Delta S^{\pm}(t) \propto \Lambda\, \left(\frac{t}{\mathcal{J}}\right)\, .
\end{align}

Then, according to (\ref{eq:complexity1_LREE}), one might write
\begin{align}
\Delta\, \mathcal{C}_{\rm cMERA}^{(1)}(t)\approx\frac{{\rm Vol}}{2}\, \Delta S_A(t)=\frac{{\rm Vol}\cdot \Lambda}{2}\, \Big(\frac{t}{\mathcal{J}}\Big) \, ,
\end{align}

which amounts to the constant growth rate given by
\begin{align}
\frac{d\, \, \mathcal{C}_{\rm cMERA}^{(1)}(t)}{dt}\approx\frac{{\rm Vol}\cdot \Lambda}{2}\, \Delta m \, .
\end{align}

Thus, over some range of time, the complexity of a state evolving after a quantum quench, measured both by $\mathcal{C}_{\mathcal{A}}$ and $\mathcal{C}_{\rm cMERA}^{(1)}$,  does increase linearly (i.e with constant complexity growth rate) and then saturates at a final value. These results remarkably show that the cMERA circuit representation (Eq.(\ref{eq:cMERA_time})) of the Hamiltonian evolution caused by a quantum quench, is able to correctly capture  growth features of both the entanglement entropy and the quantum complexity of a time-evolving state \cite{susskind16,carmimyers17}.

\section{Discussion and Conclusions}
\label{discussion}
Using a coherent state path integral representation for the cMERA circuit of a one dimensional free scalar theory, it has been shown that the variationally optimized cMERA flows extremize the action functional appearing in the path integral. The extremal values of the action functional  account for the complexity of the cMERA circuit and we suggest to interpret the path integral as a complexity functional.  Remarkably, this cMERA complexity action functional yields analogous results to the Liouville action of 2D Einstein gravity. In this interpretation, the Liouville mode is mapped to the variational parameter defining the cMERA circuit. We stress here that neither assumptions on the CA conjecture nor facts about AdS/CFT and holography/Tensor Network duality have been invoked. 

We have also introduced a kind of complexity growth limits (CGL) that constrain the entanglement renormalization flow in cMERA.  In particular, we have derived CGL of  Margolus-Levitin and Mandelstam-Tamm type that set an upper bound to the complexity growth rate in terms of the  mean value and average variance of the entanglement Hamiltonian along the renormalization group direction, respectively.  Remarkably, these bounds are saturated, a feature shared with holographic systems \cite{brown161, brown162}.  We hope that CGL will constitute a new tool that can be used to further elucidate the proposed connection between tensor networks and holography.

In this sense, it has been shown how the cMERA formulation for time dependent settings such as quantum quenches, in conjuction with the proposed action functionals that measure $\mathcal{C}_{\mathcal{A}}$, yields constant complexity rates for these time dependent cMERA RG-flows. However, in spite of cMERA being able to address these kind of time dependent states, it tackles space and time on quite different grounds. Namely, a fully covariant formulation of cMERA  is not known at this moment. As a result, we expect that a connection between time dependent versions of cMERA with recognizable gravitational theories in the bulk of the tensor network will not be so straightforward to establish. 

Finally, it would be worth to better understand  what kind of universal data can be extracted from complexity \cite{vidal17b}. One hint may come from the established connection between the complexity measure based on the cMERA circuit length  and the amount of Left-Right Entanglement created along the RG-flow. Further insight may be obtained  from the connection between the cMERA variational parameter and the entanglement entropy of an interval of the cMERA target state. It is worth investigating which approach, if any,  could lead to additional conformal data.

\acknowledgments
It is a pleasure to thank M.T. Mueller for insightful and helpful discussions. We acknowledge funding support from Spanish Ministerio de Econom\'{i}a y Competitividad (project FIS2015-69512-R), Programa de Excelencia de la Fundaci\'{o}n S\'{e}neca Regi\'{o}n de Murcia (project 19882/GERM/15), the John Templeton Foundation, and UMass Boston (project P20150000029279).

\end{document}